\documentclass[runningheads]{cl2emult}

\usepackage{makeidx}  
\usepackage{graphicx} 
\usepackage{subeqnar} 
\usepackage{multicol} 
\usepackage{cropmark} 
\usepackage{eso}      
\makeindex            

\def\etal{et al.}
\def\lya{Ly$\alpha$ }
\def\kms{km  s$^{-1}$}
\def\ciis{C II$^{*}$ }
\def\micron{$\mu$m }
\def\perd{\;\;\; .}
\def\cmma{\;\;\; ,}
\def\lclos{$l_{c}$}
\def\lcav{$<l_{c}>$}
\def\lcr{$l_{cr}({\rm {\bf r}})$}
\def\gamd{$\Gamma_{d}({\rm {\bf r}})$}
\def\nh{$N$(H I)}

\begin{document}
\title*{Measuring Feedback in Damped {\lya} Systems}
%
%
%
%
\titlerunning{Feedback}
%
\author {Arthur~M.~Wolfe}
\authorrunning{A. M. Wolfe}
%
%
\institute{University of California, La Jolla CA 92093-0424, USA}

\maketitle              

\begin{abstract}

We measure feedback (heating rates) in damped {\lya} systems 
(DLAs)
from the cooling rate of the neutral gas. 
Since cooling occurs
through [C II] 158 {\micron} emission,  we infer cooling
from {\ciis} 1335.7 absorption lines detected with HIRES on the
Keck I telescope. The inferred heating rates
are 30 times lower than for the Galaxy ISM. At $z$ $\approx$
2.8 the implied
star formation rate per unit area is log$\psi_{*}$=$-$2.4$\pm$0.3
$M_{\odot}$ y$^{-1}$ kpc$^{-2}$ and the 
star formation rate per unit comoving
volume log$d{\rho_{*}}/dt$=$-$0.8$\pm$0.2$M_{\odot}$ y$^{-1}$ Mpc$^{-3}$. 
This is
the first measurement of star formation rates in objects 
that typify the protogalactic mass distribution.

\end{abstract}

\section{Introduction}

The purpose of this paper is
to discuss a new method for probing physical conditions in  DLAs.
In principle, this technique can be used to probe their masses, 
sizes, and baryonic content. The focus of this paper is on
star formation. Specifically, 
in this paper I infer the star formation rates per unit area in 
DLAs by 
measuring the rate
at which the {\em neutral} gas is
heated. This is possible because it is reasonable to 
attribute heating to far UV radiation emitted by massive
stars.  
Since the cross-sectional area times the comoving
density of the DLAs is known,
we can derive the rate of star formation per unit comoving
volume. Until now, comoving star formation rates
have been obtained for highly luminous objects
such as the Lyman Break Galaxies. Here I derive comoving star formation rates for
objects that are more representative of the protogalactic mass distribution.

\section{[C II] 158 {\micron} Emission From Damped {\lya} Systems}

[C II] 158 {\micron} emission results from transitions between the $^{2}P_{3/2}$
and $^{2}P_{1/2}$ fine-structure states in the ground term of C$^{+}$. It dominates
the cooling rate by the Galaxy ISM with a luminosity $L$([C II]) = 5$\times$10$^{7}$$L_{\odot}$
~\cite{ref2.1}. Most of the emission from the Galaxy and other nearby spirals arises
in the diffuse neutral gas rather than from star-forming regions in spiral arms, or
PDRs on the surfaces of molecular clouds (e.g. ~\cite{ref2.2}). The last point is
especially relevant for DLAs where molecules are rarely detected ~\cite{ref2.3}.

In order to estimate the [C II] 158 {\micron} emission rates, Pottasch {\etal} ~\cite{ref2.4}
derived the following expression for the 158 {\micron} luminosity per H atom:

\begin{equation}
l_{c} = N({\rm C II}^{*})h{\nu}_{ul}A_{ul}/N({\rm H I}) \ {\rm erg \ s^{-1} \ H^{-1}}
\cmma
\end{equation}

\noindent where  $N$({\ciis}) is the column density of C$^{+}$ ions in the $^{2}P_{3/2}$
state, {\nh} is the H I column density, and $A_{ul}$ and $h{\nu}_{ul}$ are the Einstein
$A$ coefficient and energy of the $^{2}P_{3/2}$ $\rightarrow$ $^{2}P_{1/2}$ transition.
In fact, {\lclos} is the density$-$weighted average along the sightline of 
{\lcr}, the emission rate per H atom at a given displacement
vector ${\rm {\bf r}}$. Here
$l_{cr}({\rm {\bf r}}) = {{n_{{\rm CII}^{*}}({\rm {\bf r}})A_{ul}h{\nu}_{ul}} / { n_{{\rm HI}}
({\rm {\bf r}})}}$,
where $n_{{\rm CII}^{*}}$ and $n_{{\rm H I}}$ are the volume densities of H I and {\ciis}.

Figure 1 shows estimates of {\lclos} versus {\nh}, where $N$({\ciis})
is deduced from the strength of {\ciis} 1335.708 absorption and {\nh} from DLA
absorption. The small stars depict measurements for several sightlines
through the ISM \cite{ref2.4} and \cite{ref2.5}. 
The large star
depicts {\lcav}, i.e., {\lcr} averaged over the H I disk of the Galaxy. 
Figure 1 also shows our measurements of {\lclos} for 27 DLAs.
As neither 158 {\micron} nor 21 cm emission has been detected from any
DLAs, {\lcav} cannot be inferred directly, where
in this case {\lcav} corresponds to {\lcr} integrated over
the volume of the average DLA. Nevertheless, comparison
between the two distributions of {\lclos} clearly indicates {\lcav} for the
DLAs to be a factor of 30 or more lower than for the ISM.

\begin{figure}
\centering
\includegraphics[width=.57\textwidth]{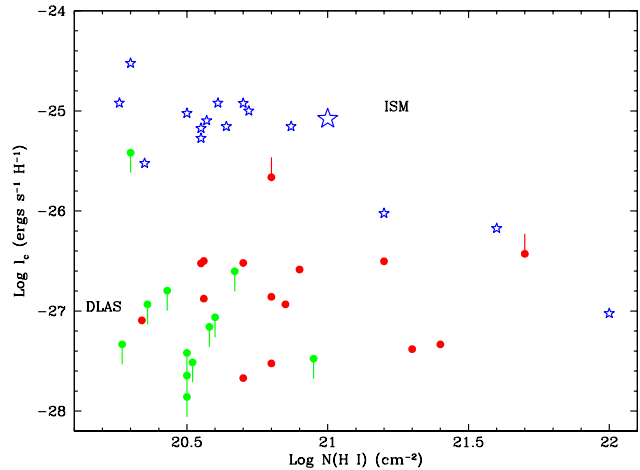}
\caption[]{$l_{c}$ versus {\nh} for DLAs (solid circles) and sightlines
through the ISM (stars). Upper limits are 2-{$\sigma$}. Large star is {\lcav}=$L$([C II])/(${m_{H}}M$(H I)) where $L$([C II]) and $M$(H I) 
are the 158 {\micron} luminosity and H I mass of the Galaxy.}
\label{eps1}
\end{figure}

\section{PROPERTIES OF MULTI-PHASE MEDIA}
\label{mlphs}

\begin{figure}
\centering
\includegraphics[width=.45\textwidth]{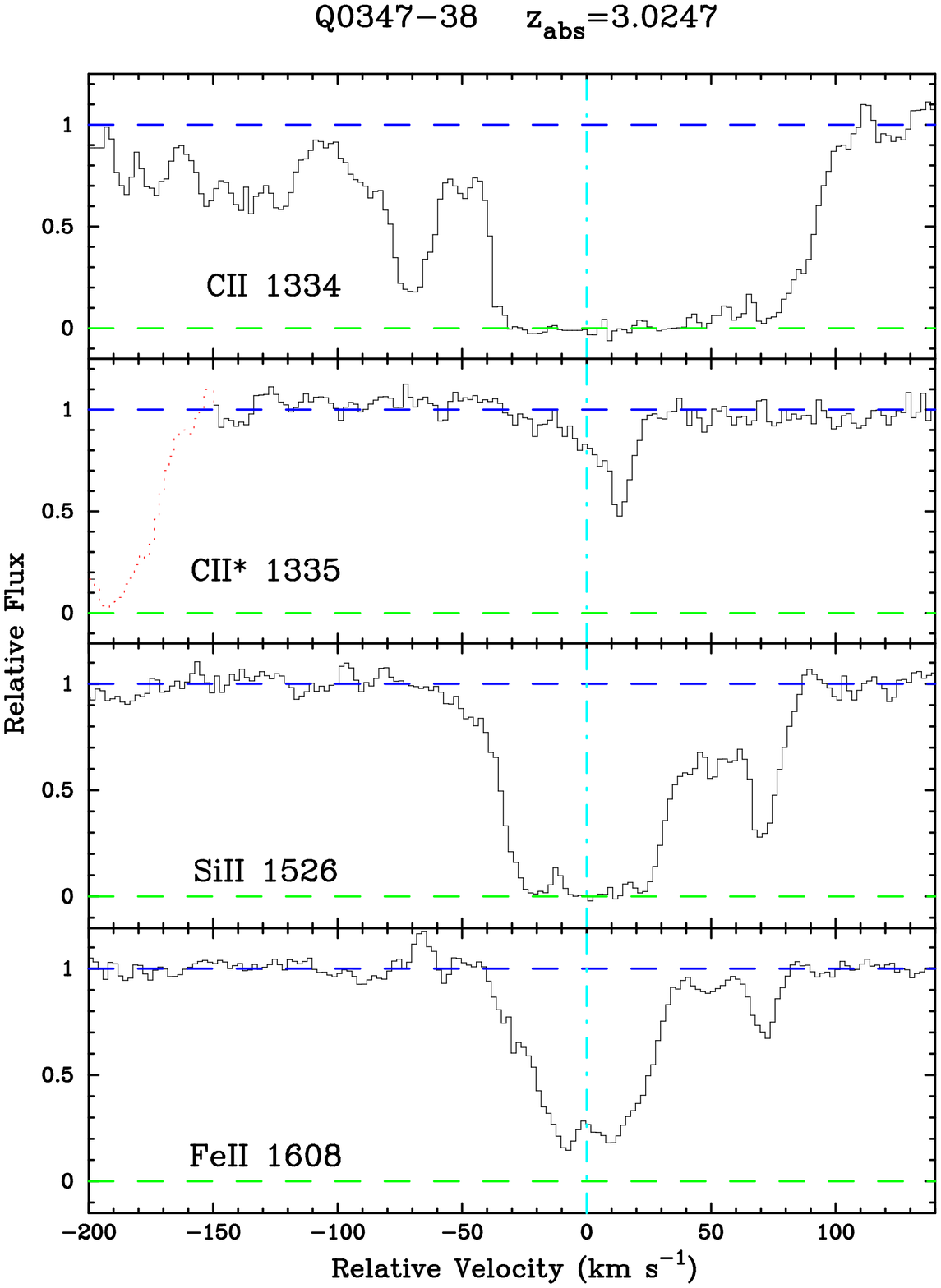}
\caption[]{Keck HIRES absorption profiles for {\ciis} 1335.7 and 3 resonance lines in the
$z$ = 3.0247 damped absorber toward Q0347$-$348. Comparison between the
{\ciis} and unsaturated Fe  II 1608 profiles shows evidence for a CNM
at $v$ = $+$ 10 {\kms} and WNM at $-$ 10 {\kms}}
\label{eps2}
\end{figure}

To determine the heating rate from measurements of {\lclos} we consider the
thermal and ionization equilibria of gas subjected to heating and cooling. The
result  is a multi-phase
medium in which the various phases are in pressure equilibrium. There are
indications that the H I gas
in DLAs is in
two phases: a cold ($T \ \ge$ 50 K) neutral medium (CNM), and a warm ($T \ \sim$
8000K) neutral medium (WNM).
The detection of 21
cm absorption in DLAs indicates the presence of a CNM since
the 21 cm optical depth is inversely proportional to temperature ~\cite{ref3.1}.
Evidence for a WNM comes from misalignment in
velocity space
between 21 cm absorption or  {\ciis} 1335.7 absorption on the one hand and
UV resonance lines  on the other 
(see Figure 2), since the CNM dominates 158 {\micron} emission (Fig. 3c).

To compute the properties of the multi-phase gas we focus on the thermal and
ionization balance of the atomic CNM and WNM.  By analogy with
the ISM we assume that heating is dominated
by photoelectric ejection of electrons from grains illuminated by far UV radiation
($h{\nu}$=8$\rightarrow$13.6 eV). 
We adopt the formalism of ~\cite{ref3.2}
who considered small grains and PAHs
and found the photoelectric heating rate at ${\rm {\bf r}}$ to be given by

\begin{equation}
{\Gamma}_{d}({\rm {\bf r}}) = 1.0{\times}10^{-24}{\epsilon}G_{0}({\rm {\bf r}}) \ {\rm ergs \ s^{-1} \ H^{-1}}
\perd
\end{equation}

\noindent In the last equation $G_{0}({\rm {\bf r}})$, the radiation field incident
at {\rm {\bf r}}, equals 4$\pi$$J({\rm {\bf r}}$)
and is in units of Habing's ~\cite{ref3.3} estimate of
the local interstellar value (=1.6{$\times$}10$^{-3}$ ergs cm$^{-2}$ s$^{-1}$).
Here $J$ is the mean intensity integrated between 8 and 13.6 eV,
and {$\epsilon$} is the
far-UV heating efficiency. 
Following ~\cite{ref3.2}
we include ionization and heating by cosmic rays 
and assume 
that the dust-to-gas ratio equals the metallicity, 
$Z/Z_{\odot}$. As a result {\gamd} is proportional to $Z/Z_{\odot}$.

An example of the computed 2-phase diagrams is shown in Figure 3.
Here
we assume $G_{0}$ = 1.7 which corresponds to the far UV radiation 
field in the ISM (see ~\cite{ref3.2}). We also let the
cosmic ray ionization rate $\zeta$ = 1.8$\times$10$^{-17}$ $s^{-1}$ 
(see ~\cite{ref3.2}). 
For comparisons with the DLA data 
we let 
[Fe/H] ($\equiv$ log$Z/Z_{\odot}$) = $-$1.5, 
the mean metallicity of the DLAs
(~\cite{ref3.4}).
Since regions of thermal stability
occur where
${\partial}({\rm log} P)/{\partial}({\rm log}n)$ $>$ 0,
a stable 2-phase medium, comprising the WNM and CNM,
can exist between the local pressure
minimum and maximum. 
Figure 3c shows 
that {\lcr} asymptotically approaches a maximum at log $n$ $>$ 1.5 cm$^{-3}$ in 
the CNM where it dominates the cooling rate.
Comparison with Figure 1 further indicates 
that this {\lcr} is 
much lower than  the {\lcav} observed in
the ISM, but falls within the range of {\lclos} spanned by the DLAs.

\begin{figure}
\centering \includegraphics[width=.55\textwidth]{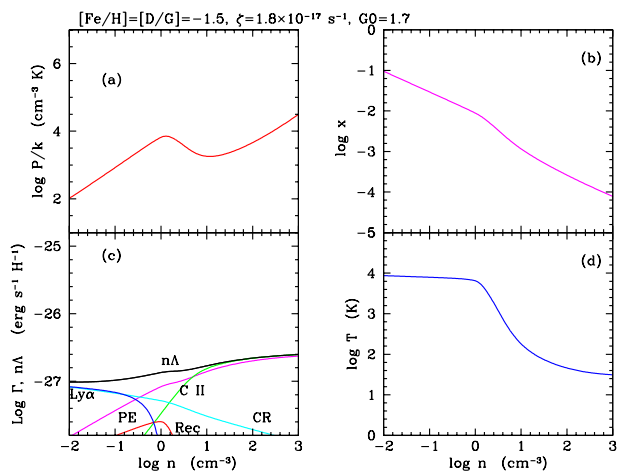}
\caption[]{(a)Pressure vs. H density, $n$. (b) Electron fraction vs. $n$. (c) Photoelectric
(PE) and cosmic-ray (CR) heating rates vs. $n$.  {\lya}, [C II] (i.e., {\lcr}), 
grain recombination (Rec), and
total ($n{\Lambda}$) cooling rates vs. $n$. (d) Temperature vs. $n$.}   
\label{eps3}
\end{figure}

\section{Star Formation Rates} 

\begin{figure}
\centering \includegraphics[width=.45\textwidth]{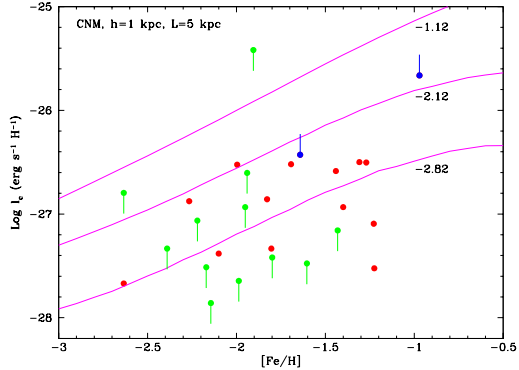}
\caption[]{Computed {\lcr} vs. [Fe/H] for log$\psi_{*}$ = -2.8,-2.1,
and -1.1 $M_{\odot}y^{-1}kpc^{-2}$. Data for DLAs
show {\lclos} versus [Fe/H].} 
\label{eps4}
\end{figure}

The previous discussion suggests
that the far UV radiation field in high-$z$ DLAs
is similar to that in the  Galaxy. 
Because the far UV intensity
is proportional to the rate at which massive stars form 
~\cite{ref4.1},
the star formation rates  
should be likewise
similar. More precisely, the similarity is between the star formation rates
per unit area, since solutions to the radiative transfer equation for
plane parallel slabs of dust and stars show that  
$J_{\nu}$ is proportional to the luminosity per unit area, $\mu_{\nu}$
~\cite{ref4.2}. Here we compute $J_{\nu}$ at the center of a 
uniform slab of stars
and dust, and then
integrate solutions for $J_{\nu}$ to 
$J$ with ~\cite{ref4.2}'s fit to the far UV spectrum of the ISM
radiation field. Finally we combine $J$ 
with eq. (2) to compute the heating rates.

The curves in Figure 4 plot the resultant {\lcr} versus [Fe/H]. 
The {\lcr} are computed  in the CNM limit,
{\lcr}={\gamd}, and for {\bf r} corresponding to the central points of
uniform slabs. The curves are parameterized by 3 different values of $\psi_{*}$,
the star formation rate per unit area. Comparison with the DLA data
reveals 
significant variations of {\lclos} at fixed
[Fe/H]. This indicates that
$\psi_{*}$ is distributed at constant metallicity. The data also show tentative evidence for
the predicted decrease in {\lcr} with decreasing [Fe/H] 
at an apparent upper limit to the star formation rate of log$\psi_{*}{\approx}-2.12$
$M_{\odot}$y$^{-1}$kpc$^{-2}$. However, we cannot rule out entirely a possibility
currently under investigation; namely, 
that the DLA sightlines pass through gas containing only WNM ~\cite{ref4.3} in which
case the $\psi_{*}$ in Figure 4 are lower limits.

\begin{figure}
\centering \includegraphics[width=.4\textwidth]{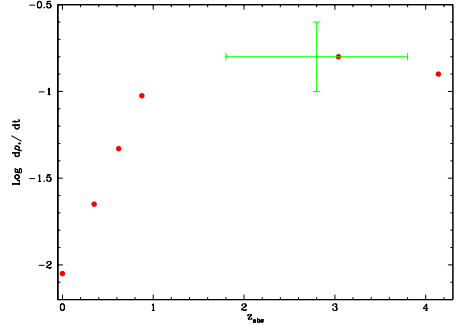}
\caption[]{Star formation rates per unit comoving volume computed
for $\Omega_{m}=1$,$\Omega_{\Lambda}=0$,$h$=0.5 cosmology.
Circles denote
luminous galaxies {\cite{ref4.5}} and point with error bars, 
DLAs. Horizontal error bar depicts redshift range of DLA sample. }
\label{eps5}
\end{figure}

The star formation rate per unit comoving volume is given by

\begin{equation}
d{\rho_{*}}/dt= {{<{\psi_{*}}>(A/A_{p}){\times}dN/dz} \over {{(1+z)^{3}}(-cdt/dz)}} 
\perd
\end{equation}

\noindent where 
$<{\psi_{*}}>$ is the average star formation
rate per unit area, $A$ and $A_{p}$ are the average  cross-sectional area and cross-sectional
area projected on the sky, 
and
$dN/dz$ is the number of DLAs per unit redshift interval.
Assuming log$<{\psi_{*}}>=-2.4{\pm}0.3$ $M_{\odot} \ y^{-1} \ kpc^{-2}$
and $dN/dz$=0.24 at $z$ = 2.8
~\cite{ref4.4}, 
the median redshift of our sample,
we
find that
in the plane-parallel limit (i.e., $A=2A_{p}$), log$d{\rho_{*}}/dt$
=$-$0.8$\pm$0.2 $M_{\odot}y^{-1}Mpc^{-3}$.
This result, plotted in Figure
5, is compared with estimates derived from objects detected in emission
such as the Lyman Break galaxies
~\cite{ref4.5}. It is encouraging that the DLA measurement is not widely different
from measurements made with independent techniques. 
The similarity  could mean that the DLAs and Lyman Break galaxies
are the same entity.  More likely, our estimates of $\psi_{*}$
and $d{\rho_{*}}/dt$
suggest the DLAs to be characterized by low
specific star formation rates and high comoving densities relative to the Lyman Break
galaxies.

{\bf Acknowledgements:} I wish to thank C. F. Mckee and J. X. Prochaska for many valuable comments.
This research was partially supported by NSF grant AST0071257.

\clearpage
\addcontentsline{toc}{section}{Index}
\flushbottom
\printindex


\begin{thebibliography}{7}
%
\addcontentsline{toc}{section}{References}


\bibitem{ref2.1} Wright, E. L. {\etal} {1991} Preliminary Spectral observations
of the galaxy with a 7$^{'}$ beam by the cosmic background explorer (COBE), 
Ap. J., {\bf 381}, 200--209


\bibitem{ref2.2} Madden, S. C. {\etal} {1993} 158 micron [C II] mapping of NGC 6946:
probing the atomic medium,
Ap. J., {\bf 407}, 579--587

\bibitem{ref2.3} Lu, L., Sargent, W. L. W., \& Barlow, T. A. (1998),
Abundances of Heavy elements and CO molecules in
high redshift damped {\lya} Galaxies,
in Highly Redshifted Radio Lines, ASP Conf. Ser. 156, 
 eds. C. L. Carilli {\etal}, 132--137 
243--250.


\bibitem{ref2.4} Pottasch, S. R., Wesselius, P. R., \& van Duinen, R. J.  (1979)
Determination of cooling rates in the interstellar medium,
A\&A, {\bf 74}, L15--L17

\bibitem{ref2.5} Gry, C., Lequeux, J.,  \& Boulanger, F. (1992)
Fine structure lines of C$^{+}$ and N$^{+}$ in the galaxy
A\&A, {\bf 266}, 457--462


\bibitem{ref3.1} Chengalur, J. N., \& Kanekar, N. {2000}, Implications
of 21 cm observations for damped {\lya} systems,
astro-ph/00115401
 
\bibitem{ref3.2} Wolfire {\etal} (1995) The neutral atomic phases of the
interstellar medium,
Ap. J., {\bf 443}, 152--168

\bibitem{ref3.3} Habing, H. J.  (1968) The interstellar radiation
density between 912 {\AA} and 2400 {\AA},
Bull. Astr. Inst. Neth.  {\bf 19}, 421--431


\bibitem{ref3.4} Prochaska, J. X. \& Wolfe, A. M. (2000) Metallicity evolution
in the early universe
ApJ, {\bf 533}, L5--L8

\bibitem{ref4.1} Madau, P., Pozzetti, L., \& Dickinson, M. (1998) The star formation 
history of field galaxies
ApJ, {\bf 498}, 106--116

\bibitem{ref4.2} Draine, B. T.  (1978) Photoelectric heating of the interstellar gas
ApJS, {\bf 35}, 595--619

\bibitem{ref4.3} Norman, C. A. \& Spaans, M.  (1997) Molecules at high
redshift: the evolution of the cool phase of protogalactic disks
Ap. J., {\bf 480}, 145--154
 
\bibitem{ref4.4} Storrie-Lombardi L., \& Wolfe, A. M.  (2000) Surveys for
$z \ >$ 3 Damped {\lya} Absorption Systems: The Evolution of the Neutral
Gas. Ap. J., {\bf 543}, 552--576

\bibitem{ref4.5} Pettini, M. (2000) Private communication
	



\end{thebibliography}
\end{document}